# Magnetic and universal magnetocaloric behavior of rare-earth substituted DyFe$_{0.5}$Cr$_{0.5}$O$_3$

Mohit K. Sharma and K. Mukherjee*

School of Basic Sciences, Indian Institute of Technology Mandi, Mandi Himachal Pradesh India 175005

*Email:kaustav@iitmandi.ac.in

## ABSTRACT

We report the effect of partial substitution of Dy-site by rare-earths (*R*=Gd, Er and La)on the magnetic and magnetocaloric behavior of a mixed metal oxide DyFe$_{0.5}$Cr$_{0.5}$O$_3$.Structural studies reveal that substitution of Dy by *R* has a minimal influence on the crystal structure. Magnetic and heat capacity studies show that the magnetic transition around 121 K observed for DyFe$_{0.5}$Cr$_{0.5}$O$_3$ remains unchanged with rare-earth substitution, whereas the lower magnetic transition temperature is suppressed/enhanced by magnetic/non-magnetic substitution. In all these compounds, the second order nature of magnetic transition is confirmed by Arrott plots. As compared to DyFe$_{0.5}$Cr$_{0.5}$O$_3$, the values of magnetic entropy change and relative cooling power are increased with magnetic rare-earth substitution while it decreases with non-magnetic rare-earth substitution. In all these compounds, magnetic entropy change follows the power law dependence of magnetic field and the value of the exponent *n* indicate the presence of ferromagnetic correlation in an antiferromagnetic state. A phenomenological universal master curve is also constructed for all the compounds by normalizing the entropy change with rescaled temperature using a single reference temperature. This master curve also reiterates the second order nature of the magnetic phase transition in such mixed metal oxides.

## 1. Introduction

Investigation of materials showing large magnetocaloric effect (MCE) has been an important area of research for its potential application in magnetic refrigeration technology [1-4]. Magnetic refrigeration provides a highly efficient and environment friendly cooling in comparison to conventional gas compression/expansion techniques [3, 4]. Generally, in two temperature ($T$) regimes this technology is important: near the room $T$ where it can be used for domestic and industrial refrigeration, and also, in the low $T$ where it can be useful for some specific technological applications in liquefaction of hydrogen in fuel industry and space science [1-3, 5 and 6]. For a material to be a good solid refrigerant it should have a high density of magnetic moments and a strong $T$ and field dependence of the magnetization for the occurrence of a magnetic phase transformation around the working $T$. Additionally, the material should have insignificant magnetic hysteresis to avoid energy losses during the magnetization/demagnetization cycles [3, 4]. Also the material should have large magnetic entropy change ($\Delta S_M$) and large relative cooling power (RCP). Apart for technological applications, from the viewpoint of basic physics, investigation of magnetocaloric parameters of magnetic materials is interesting and important as one can acquire insight about complex magnetic phases present in the system, which may not be possible by just studying magnetization. For example, a detailed analysis of the field dependence of magnetocaloric effect can provide useful information about the performance of a refrigerant for magnetic field ranges used in actual refrigeration cycles. Beside this, such a study can also be helpful to get deeper understanding of the nature of magnetic phase transitions and phase coexistence in the material.

Many oxides containing rare-earths and transition metals have been found to be showing good MCE [4]. In this context rare-earth transition metal perovskites of the form $ABO_3$ (where A is the rare-earth ion and B is the transition metal ion) which show magnetic transition at low $T$ are interesting for investigation of low $T$ MCE [7-12]. In last few years, a new family of mixed metal oxides (combining orthoferrites and orthochromites) of the form $RFe_{0.5}Cr_{0.5}O_3$, is discovered in perovskites and are being extensively investigated [13-18]. Such compounds are important as combining the two transition metals within the perovskite structure can be an effective approach to enhance the magnetic properties and at the same time tune/induce functional properties as compared to their parent compounds. For example, a single compound, $DyFe_{0.5}Cr_{0.5}O_3$, exhibits both magnetoelectric (ME) coupling as well as MCE [13 and16], whereas $DyCrO_3$ shows only MCE [9] while $DyFeO_3$ shows magnetic field induced multiferroicity [19]. DyFe$_{0.5}$Cr$_{0.5}$O$_3$ shows complex magnetic nature due to co-existence of several types of order parameters. In such compounds magnetic transition *T* and hence the MCE can be tuned by substitution at the Dy-site.

Hence, in this paper we report an extensive investigation magnetic and magnetocaloric properties of a series of compounds Dy$_{0.8}$R$_{0.2}$Fe$_{0.5}$Cr$_{0.5}$O$_3$ (*R* = Gd, Er and La). DyFe$_{0.5}$Cr$_{0.5}$O$_3$ undergoes in three antiferromagnetic (AFM) ordering around *T*~261 K, *T$_1$* ~ 121 K and *T$_2$* ~13 K [13 and 16]. It is observed that *T$_1$*is unaffected, but *T$_2$* is changed due to partial replacement of Dy. Due to weakness of magnetic features around 261 K, the effect of partial substitution on this transition could not be tracked. Hence, in this manuscript we have restricted our study in the *T* range of 2 to 150 K. Interestingly a significant MCE is observed only around the *T$_2$* in all these compounds. The MCE and RCP values are increased/decreased with magnetic/nonmagnetic substitution with respect to that observed in DyFe$_{0.5}$Cr$_{0.5}$O$_3$. Arrott plot confirms the second order nature of the magnetic transition in all the compounds and magnetic entropy change follows the power law of the dependence of magnetic field of the form $\Delta S_M \sim H^n$. The obtained value of exponent *n* indicates the presence of ferromagnetic (FM) correlation in an AFM state in all these compounds. A phenomenological universal curve of all the compounds is created by normalizing the entropy change with rescaled *T*. This master curve also restates the second order nature of the magnetic phase transition in such materials.

2. **Experimental**

Polycrystalline samples of Dy$_{0.8}$Gd$_{0.2}$Fe$_{0.5}$Cr$_{0.5}$O$_3$ (DGFCO), Dy$_{0.8}$Er$_{0.2}$Fe$_{0.5}$Cr$_{0.5}$O$_3$ (DEFCO) and Dy$_{0.8}$La$_{0.2}$Fe$_{0.5}$Cr$_{0.5}$O$_3$ (DLFCO) were prepared by solid state reaction method under the similar conditions as reported in Ref [16]. The DyFe$_{0.5}$Cr$_{0.5}$O$_3$ (DFCO) sample is the same as used Ref [16]. The structural analysis is carried out by x-ray diffraction (XRD, Cu K$\alpha$) using Rigaku Smart Lab instruments. The Rietveld refinement of the powder diffraction data is performed using FullProf Suite software. Temperature (2 - 300 K) and magnetic field (*H*) (up to 50 kOe) dependent magnetization were performed using Magnetic Property Measurement System (MPMS) from Quantum design, USA. Heat capacity (*C*) measurements in the *T* range 2-150 K were performed using the Physical Property Measurements System (PPMS).

## 3. Results and discussion

Fig. 1 show the XRD patterns of all the compounds recorded at room $T$. All these compounds crystallize in orthorhombic perovskite structure with Pbnm space group. The crystallographic parameters obtained from Rietveld analysis are listed in Table 1. A small shifting in peak with respect to parent compound is observed (inset Fig. 1), which confirms the expansion and contraction of lattice with $R$ substitution; implying that the dopant goes to the respective sites.

Fig. 2 (a)-(c) shows the $T$ response of dc magnetic susceptibility ($\chi$) data taken in zero field cooling (ZFC) and field cooled (FC) condition in the $T$ range from 2 to 150 K at 100 Oe for DGFCO, DEFCO and DLFCO compounds. All these compounds undergo two distinct AFM ordering as confirmed by d$\chi$/d$T$ vs $T$ plots. For all the compounds this curve shows a slope change around $T_1 \sim 121$ K (not shown) and a peak at $T_2 \sim 9$, 11 and 18 K respectively (inset of Fig. 2(b); curve for DFCO is added from Ref [16] for comparison). DFCO is reported to undergo two AFM ordering around 121 and 13 K [13, 16]. It is observed that partial replacement of Dy by $R$ ions result in shifting of the $T_2$ while $T_1$ remains unchanged. It has been reported in DFCO that the magnetic ordering at $T_1$ is caused due to the Cr-O-Cr ordering, while the ordering at $T_2$ is caused due to Dy-O-Fe/Cr magnetic interactions [13]. To confirm magnetic ordering behaviour of these compounds, heat-capacity is measured as a function of $T$ (2-150 K) shown in Fig 2 (d)-(e). For all the compounds, a weak, but a distinct peak is observed around $T_2$, but such a peak around $T_1$ is not clear in the raw data. In order to see the features more clearly, the $T$ response of d$C$/d$T$, is plotted in the insets of Fig. 2 (e) and (f). The derivative curves show a minima and peak around $T_2$ and $T_1$ respectively. These features in $C$ measurements confirm the magnetic ordering temperatures of the respective compounds. Hence from the above observations, it can be said that partial replacement of Dy by $R$ ions does not affects the magnetic ordering due to Cr. However, with magnetic dopants $T_1$ is suppressed while with non-magnetic dopant it is enhanced. In these compounds, there is a competition between transition metal sub-lattice moments (at B-site) and rare earth sub-lattice moments (at A-site). Due to the dominance of rare earth field the ordering of moments is observed at low $T$. In fact that Gd and Er might be on the threshold of magnetic ordering as the $T$ is lowered to 2 K and the internal magnetic field arising because of this, however small it may be, is such that it suppresses AFM ordering due to Dy. However, when Dy site is diluted by nonmagnetic La ions, transition metal sub-lattice moments dominate resulting in the observation magnetic ordering at higher $T$.

To get a better understanding about the magnetic behavior and effect of substitution, isothermal $M$ is measured as a function of $H$ (in the range ±50 kOe) at different $T$ (2 to 150 K). Fig. 3 (a-c) exhibits the representative $M$ $(H)$ curves at selected temperatures of 2, 40 and 150 K for DGFCO, DEFCO and DLFCO compounds. For all these compounds, a weak magnetic hysteresisis observed below 5 kOe (shown in insets) and magnetic saturation is absent at high fields. Such behaviour indicates presence of weak FM correlation along with AFM coupling in these compounds. The origin of such FM correlation is not due to the magnetic field induced metamagnetic transition, as such effect is prevalent only at low temperatures, whereas FM correlation persist upto room temperature. In fact the magnetic behavior of such mixed metal oxide is sensitive to the nature of rare earth ions, as observed from the temperature response of coercive field, which is found to be different for different composition [16]. With reference to Fig. 2 (b) of Ref [16], it is observed that, in case of magnetic doping (Gd and Er), there is an enhancement experienced by the magnetization value while there is a decrement in magnetization value with the nonmagnetic doping.

To identify the nature of magnetic transition, $H/M$ vs. $M^2$ plots were done using the virgin curves of isothermal magnetization. According to Banerjee's criterion, this Arrott plot exhibits a negative and positive slope for first and second order nature of magnetic transition Ref [20]. Fig. 4 (a)-(d) shows the $H/M$ vs. $M^2$ plots for all the compounds at selected temperatures. A positive slope is observed around $T_2$ and $T_1$ (insets of Fig. 4 (a)) indicating second order nature of the magnetic transitions.

In order to see the effect magnetic and non-magnetic $R$ substitution on the Dy site of DFCO, magnetocaloric effect is calculated from the virgin curves of $M$ $(H)$ isotherms (in the $T$ range of 2-150 K). MCE is generally measured in terms of the change in isothermal magnetic entropy ($\Delta S_M$) produced by changes in applied magnetic field. It is to be noted here, that each isotherm is measured after cooling the respective compound from room $T$ to the measurement $T$ and $\Delta S_M$ is calculated using the following expression [21]

$$\Delta S_M = \Sigma \left[ (M_n - M_{n+1})/(T_{n+1} - T_n) \right] \Delta H_n \qquad (1)$$

where $M_n$ and $M_{n+1}$ are the magnetization values measured at field $H_n$ and $H_{n+1}$ at temperature $T_n$ and $T_{n+1}$ respectively. Fig. 5 (a) shows the $T$-dependent $\Delta S_M$ at 50 kOe applied field change from zero field. An inverse MCE is observed for all the compounds in the $T_2$ region while MCE is negligible around $T_1$. Hence it can be said that in these compounds the observed MCE is due to the magnetic entropy variation arising from strong rare earth and transition metal sub-lattice interactions. Also the broadness of the $\Delta S_M$ peak indicate the second order nature of this transition [22] which is in analogy with the Arrott plots. Interestingly, it is observed that $\Delta S_M$ increases upto ~ 14.6 J/kg-K at 50 kOe for DGFCO as compared to 10.8 J/kg-K for DFCO. Er substitution results in insignificant changes to $\Delta S_M$ value (~10.9 J/kg-K) while La substitution results in slightly smaller value ~ 9.3 J/kg-K with respect to DFCO. These compounds have a higher $\Delta S_M$ at 50 kOe in comparison to other transition metal oxides like TbMnO$_3$, DyMnO$_3$ [7], DyCrO3 [9, 10] and pure and Fe doped HoCrO$_3$ [23].

In the magnetocaloric material research another parameter, namely relative cooling power (RCP) is required to further evaluate a material for their suitability in magnetic refrigeration device. RCP is the measure of the amount of heat transfer between cold and hot reservoirs in an ideal refrigeration cycle. The RCP is defined as the product of maximum $\Delta S_M$ ($\Delta S_M^{Max}$) and full width of half maximum of the peak in $\Delta S_M$ ($\Delta T_{FWHM}$), i.e.

$$\text{RCP} = \Delta S_M^{max} \times \Delta T_{FWHM} \quad \ldots\ldots (3)$$

However, in these mixed metal oxides, $\Delta S_M$ vs. $T$ plot is not found symmetric in nature. Thus, RCP is calculated by numerically integrated the area under the $\Delta S_M$ vs $T$ curve, with $T$ limit $T_c$ (cold end) and $T_h$ (hot end) of thermodynamic cycle [10]. It is expressed as

$$RCP = -\int_{T_h}^{T_c} |\Delta S_M(T, \Delta H)| dT \quad (4)$$

The $T$ response of RCP value for the studied compounds is displayed in Fig 5 (b). The observed RCP values are comparable with another manganite and chromite system [7, 8, 10 and 23]. The RCP values are increased/decreased with magnetic/nonmagnetic substitution. Therefore, mixed metal oxides show good magnetocaloric properties in the cryogenic $T$ region.

In the field of magnetocaloric material research it is essential to compare the experimental data for different materials because all MCE parameters exhibit the applied field dependent behavior and this parameter varies for compounds to compounds. Therefore, magnetic field dependent study of $\Delta S_M$ is necessary for a better understanding of intrinsic nature of MCE [24]. The field dependent $\Delta S_M$ is described in terms of power law behavior ($\Delta S_M \sim H^n$), [12, 15, 25 and 26] where $n$ is exponent directly related to magnetic state of the materials. Magnetic field response of $\Delta S_M$ fitted well with power law behavior above the peak $T$. For all studied compound, the representative curve of $\Delta S_M$ vs $H$, at selected $T$, is plotted in Figure 6 (a)-(d). Applied field dependent $\Delta S_M$ follows the power law behavior ($\Delta S_M \sim H^n$) and observed value of the exponent ($n$) lies between 1.4 to 1.6. But for ideal AFM system, the value of $n$ should be 2 [26]. This lower value of n <2 observed in these compounds is due to the presence of FM correlations in AFM state and this observation is in analogy to the inference drawn from the $M(H)$ isotherms of these compounds.

Power laws and universal scaling have been extensively used to investigate MCE in magnetic materials [15, 25-30]. They are key tools which allow us to compare the performing properties of the materials regardless of their nature, processing or experimental conditions during measurements. A phenomenological universal curve for the field dependence of $\Delta S_M$ has been proposed in Ref [29]. Generally, if a universal curve exists, then the equivalent points of the $\Delta S_M(T)$ curves measured at different applied fields should merge into the single universal curve. Universal curve are also plotted for different compositions of material [30]. The MCE data of different materials of same universality class should fall onto the same curve, irrespective of the applied magnetic field. This curve also identifies the nature of magnetic transition in materials [31]. Franco et al., and Biswas et al., have shown the universal behavior for conventional [27-29], and inverse MCE [26]. For IMCE, in compounds where symmetrical peak is observed, rescaling in $T$ axis is not required. But in these mixed metal oxides symmetrical peak is not observed and a universal curve is not obtained. However, after rescaling the $T$ axis, a universal curve is obtained for all these compounds. It is to be noted that in these compounds, the observed MCE is observed near the magnetic transition which is of second order. Generally, in such cases the $T$ axis is rescaled to obtain the universal curve using single reference temperature [30]. The rescaled temperature ($\theta$) is defined as [30]:

$$\theta = (T-T_{pk})/(T_r-T_{pk}) \qquad (2)$$

where $T_{pk}$ is peak $T$ at $\Delta S_M^{max}$ and $T_r$ is reference $T$. $T_r$ is selected as the $T$ according to the relation $\Delta S(T_r) = \Delta S_{pk}^{max}/2$. Normalized $\Delta S_M$ as function of $\theta$ at selected fields for all the compounds are shown in Fig. 7 (a)-(d). As observed from the figure, the $\Delta S_M$ curves of the compounds merges to a single phenomenological curve which is independent of field. This behavior is distinctly different than that observed for typical first order phase transition [31] and reaffirms the second order nature magnetic phase transition. Also as observed, $\Delta S_M$ vs $T$ plot display a shifting in peak $T$ with rare earth substitution with changes in $\Delta S_M$ values. Hence, we tried to construct a universal curve for different composition at same field. In order to check whether this universal curve holds for different composition of series the rescaled curves of all the compounds is plotted for maximum applied field (50 kOe). Inset of Fig. 7(a) show that the curves for all the compounds merge on single master curve of the magnetic entropy change. Thus, this series of compound follows the universal curve of $\Delta S_M$ which is calculated using a single reference temperature.

4. Conclusion

In summary, a detailed investigation of magnetic and magnetocaloric properties of rare-earth substituted mixed metal based oxide, $Dy_{0.8}R_{0.2}Fe_{0.5}Cr_{0.5}O_3$ ($R$ = Dy, Er and La) is carried out. Our results reveal that the magnetic transition around 121 K is unaffected by rare-earth substitution, whereas the lower magnetic transition $T$ is suppressed/enhanced by magnetic/non-magnetic substitution due to modification of Dy/R-O-Fe/Cr magnetic exchange interaction. Arrott plot confirms the second order nature of the magnetic transition in all these compounds. The $\Delta S_M$ and RCP values are increased/decreased with magnetic/nonmagnetic substitution. Magnetic entropy change follows the power law dependence of magnetic field and the value of the exponent $n$ indicate the presence of ferromagnetic correlation in an AFM state in all these compound. A phenomenological universal master curve of all the compounds is constructed by normalizing the entropy change with θ. This universal curve also reaffirms the second order nature of the magnetic phase transition in such materials. Thus based on the above observations, it can be said that the mixed metal oxides show good magnetocaloric properties in the cryogenic $T$ region.

**Acknowledgement**

The authors acknowledge IIT, Mandi for financial support. The experimental facilities of Advanced Materials Research Center (AMRC), IIT Mandi are also being acknowledged.

Table 1: Lattice parameters for the compounds obtained from Rietveld refinement of XRD data. The parameter for DFCO is added from Ref [16] for comparison.

| Parameters | DFCO | DGFCO | DEFCO | DLFCO |
|---|---|---|---|---|
| $a$ (Å) | 5.2860 (1) | 5.2946 (1) | 5.2781 (0) | 5.3369 (1) |
| $b$ (Å) | 5.5606 (1) | 5.5594 (1) | 5.5573 (1) | 5.5539 (1) |
| $c$ (Å) | 7.5902 (2) | 7.5982 (1) | 7.5835 (1) | 7.6395 (2) |
| $V$ (Å$^3$) | 223.02 (8) | 223.65(8) | 222.44 (9) | 226.44 (1) |
| $\chi^2$ | 1.88 | 2.46 | 1.987 | 2.6 |

Figures

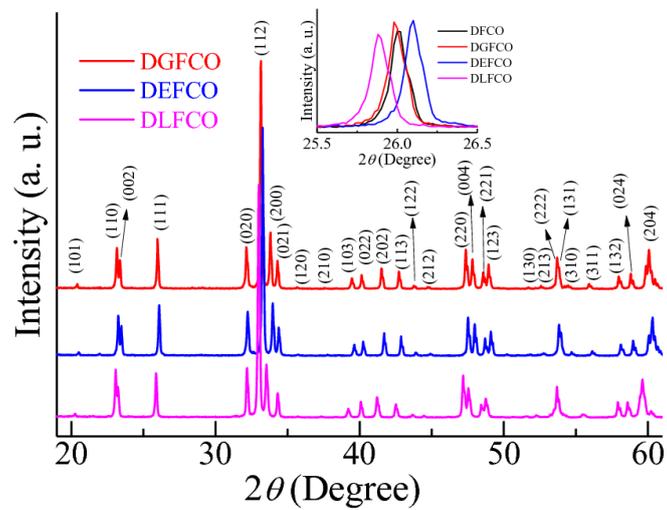

**Fig. 1**: Room temperature of X-ray diffraction patterns for DGFCO, DEFCO and DLFCO compounds. Inset: shows the expanded XRD pattern of all the compounds. Inset shows the pattern in an expanded form for one peak, to bring out that the peaks shift with substitution. The black curve is for DFCO from [16].

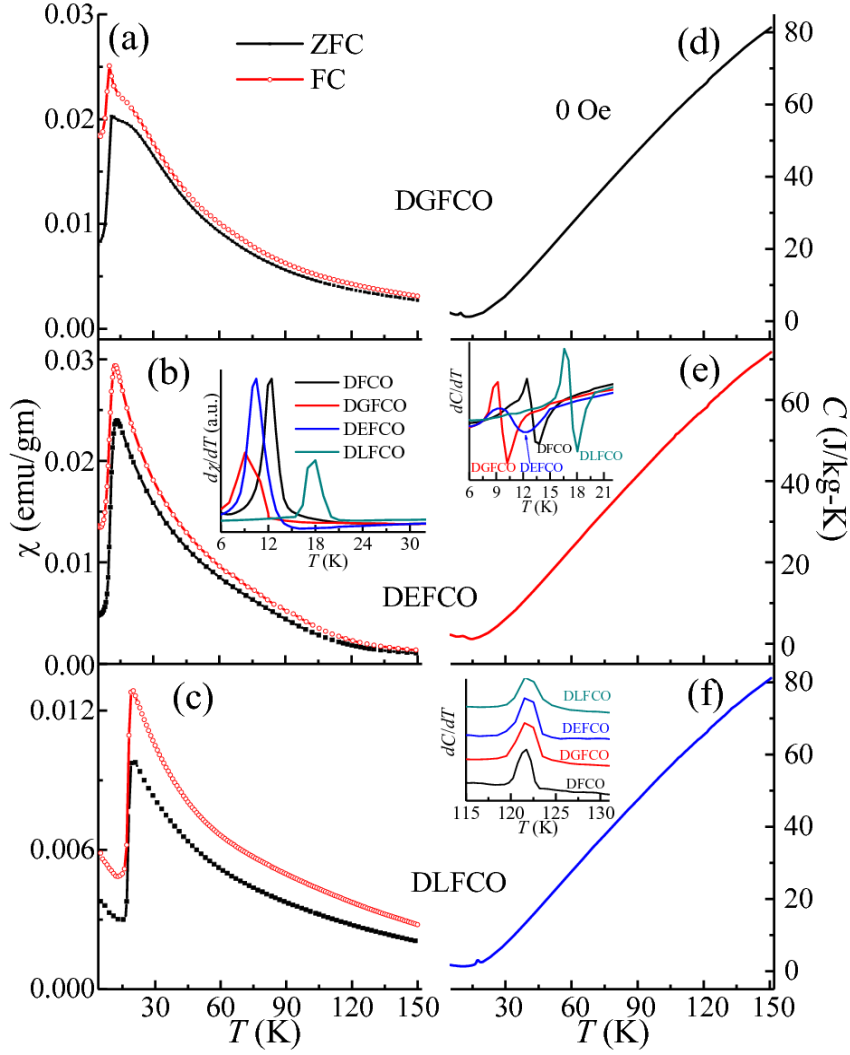

**Fig. 2**: Left Panel 1: Temperature (*T*) response of dc magnetic susceptibility (*χ* = *M*/*H*) obtained under zero-field-cooled (ZFC) and field cooled (FC) condition at 100 Oe for (a) DGFCO, (b) DEFCO and (c) DLFCO. Inset of Fig. (b): d*χ*/d*T* plotted as a function of *T* in the *T* range 2 to 30 K. The black curve is for DFCO from [16]. Right panel: *T* response of heat capacity (*C* vs *T*) for (a) DGFCO, (b) DEFCO and (c) DLFCO. Inset of Fig. (e): *dC*/*dT* vs *T* plot in the *T* range 6 - 22 K. Inset of Fig. (f): *dC*/*dT* vs *T* plot in the T range 115 - 130 K. The black curves in both the inset is for DFCO from [16].

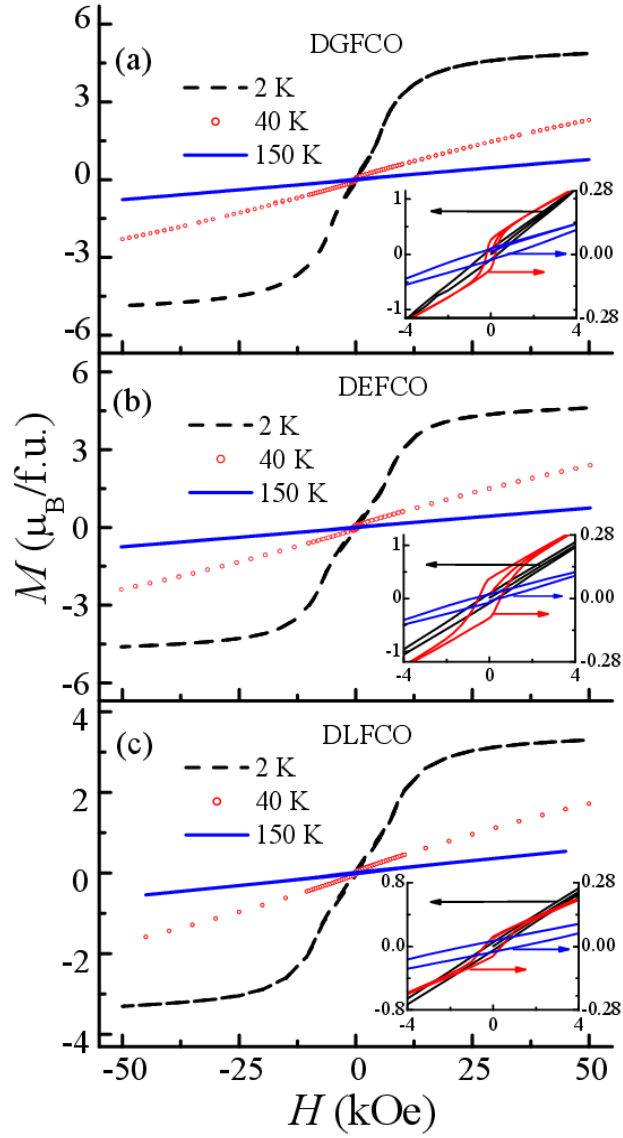

**Fig. 3**: Isothermal magnetization (*M*) plot as function of applied field (*H*) at 2, 40 and 150 K, for (a) DGFCO, (b) DEFCO and (c) DLFCO. Insets: *M* (*H*) curves expanded at the low field region for the respective compounds.

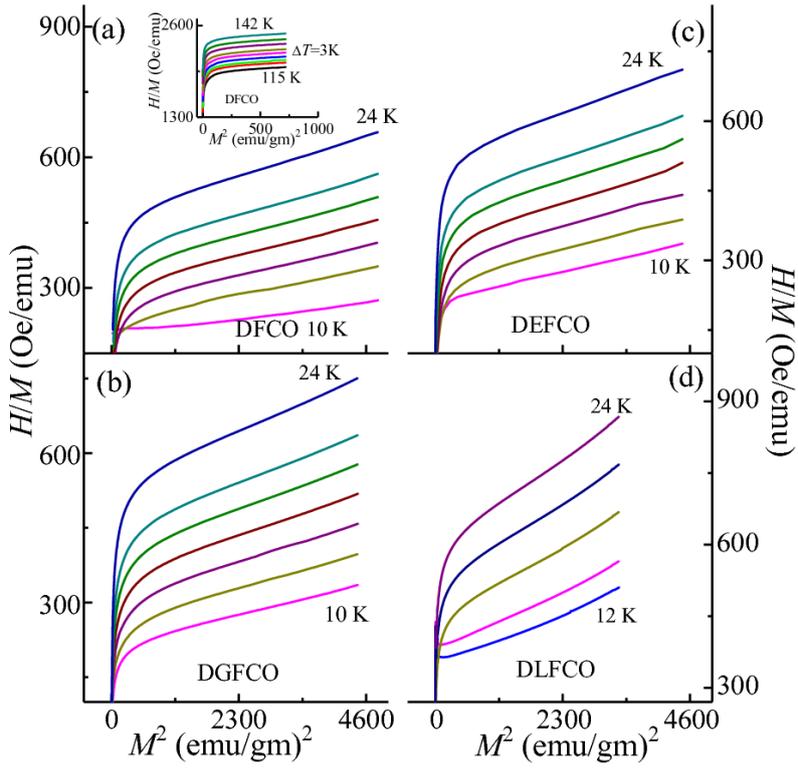

**Fig. 4**: Arrott plot ($H/M$ vs. $M^2$) for (a) DFCO, (b) DGFCO, (c) DEFCO and (d) DLFCO in the temperature range of 2-24 K. Inset: same plot in the range of 115 – 142 K for DFCO.

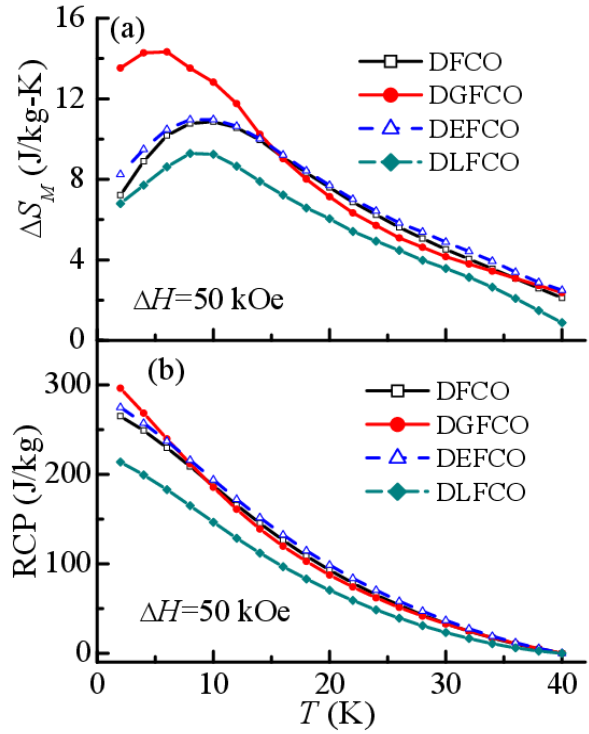

**Fig. 5:** (a) Isothermal magnetic entropy change ($\Delta S_M$) plotted as a function of temperature (*T*) for DGFCO, DEFCO and DLFCO compounds for a magnetic field change 50 kOe. The black curve is for DFCO from [16]. (b) *T* response of RCP for all the compounds.

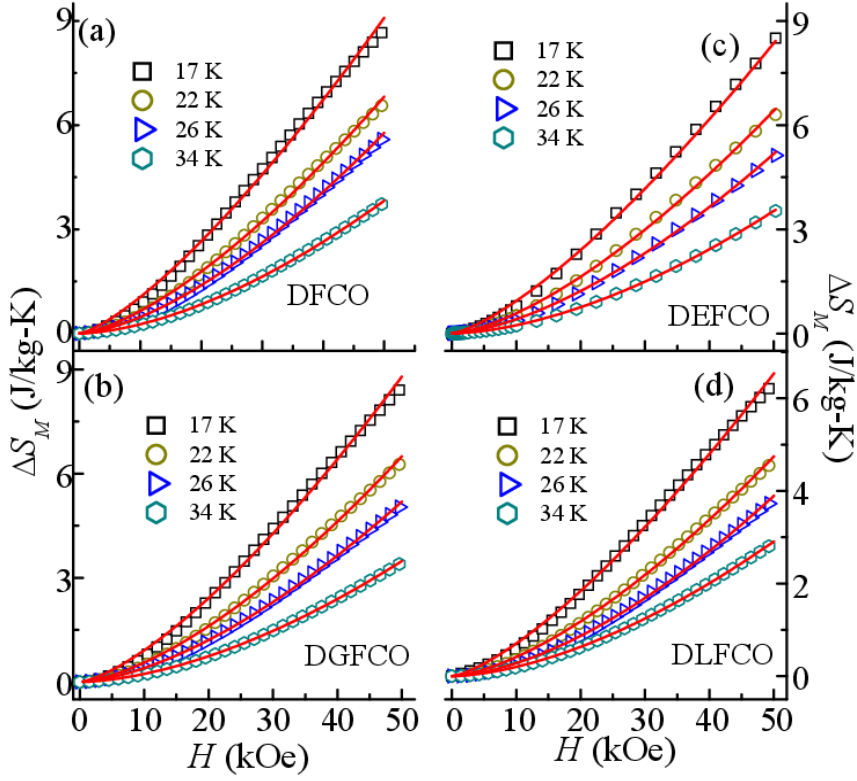

**Fig. 6**: Magnetic field (*H*) response of isothermal magnetic entropy ($\Delta S_M$) change at selected temperatures for (a) DFCO, (b) DGFCO, (c) DEFCO and (d) DLFCO. The curve through the points is the fit to the power law behaviour (as described in text).

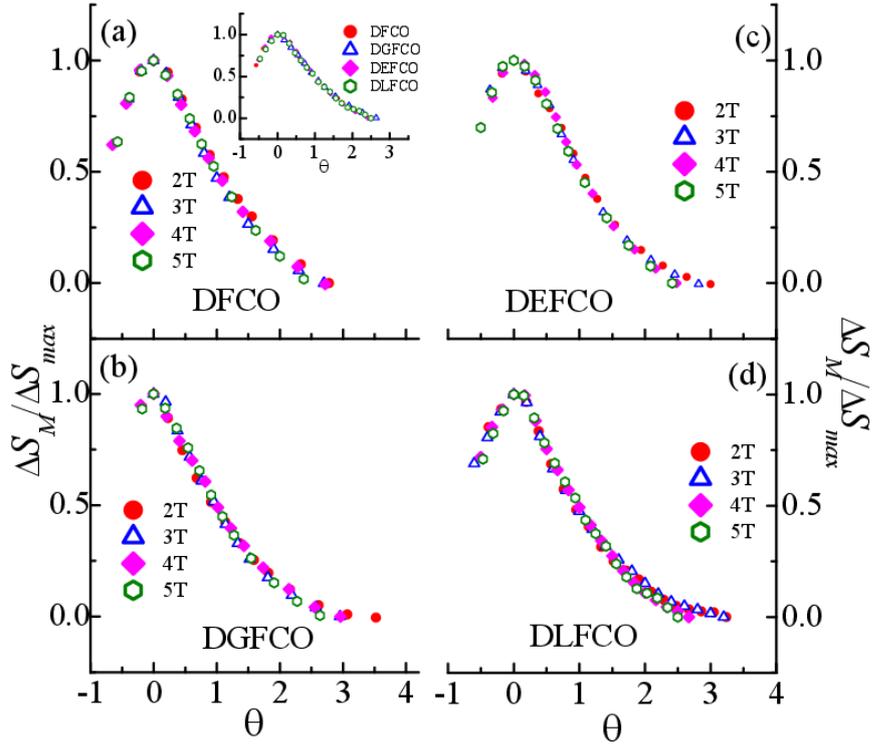

**Fig. 7**: Normalized magnetic entropy change ($\Delta S_M/\Delta S_{max}$) as a function of reduced temperature $\theta=(T-T_{pk})/(T_r-T_{pk})$ for (a) DFCO, (b) DGFCO, (c) DEFCO and (d) DLFCO. Inset of Fig (a): Same figure for all the compounds at 50 kOe.